\documentclass[aps,showpacs,twocolumn,amsmath,amssymb,nofootinbib,prc]{revtex4-1}
\usepackage{epsfig}
\usepackage{multirow}
\usepackage{amsmath}
\usepackage{amssymb}
\usepackage{epstopdf}
\usepackage{color}
\usepackage{bm}
\usepackage[bookmarksnumbered,bookmarksopen,colorlinks,citecolor=blue,linkcolor=blue] {hyperref}

\begin{document}
\title{Production cross sections for exotic nuclei with multinucleon transfer reactions}
\author{Feng-Shou Zhang$^{1,2,3,\dag}$, Cheng Li$^{1,2}$, Long Zhu$^{4}$, Peiwei Wen$^{5}$}

\affiliation{
$^{1}$The Key Laboratory of Beam Technology of Ministry of Education, College of Nuclear Science and Technology,\\
Beijing Normal University, Beijing 100875, China\\
$^{2}$Beijing Radiation Center, Beijing 100875, China\\
$^{3}$Center of Theoretical Nuclear Physics, National Laboratory of Heavy Ion Accelerator of Lanzhou, Lanzhou 730000, China\\
$^{4}$Sino-French Institute of Nuclear Engineering and Technology, Sun Yat-sen University, Zhuhai 519082, China\\
$^{5}$China Institute of Atomic Energy, Beijing 102413, China\\
Corresponding author.\ E-mail: $^{\dag}$fszhang@bnu.edu.cn
}

%\date{\today}

\begin{abstract}
The main progresses in the multinucleon transfer reactions at energies close to the Coulomb
barrier are reviewed.\ After a short presentation of the experimental progress and theoretical progress,
the predicted production cross sections for unknown neutron-rich heavy nuclei and the trans-uranium
nuclei are presented.
\end{abstract}
%\keywords{heavy-ion collisions, multinucleon transfer reactions, exotic nuclei, GRAZING model, DNS model, ImQMD model}
%\pacs{25.70.Jj, 25.70.Hi, 24.10.Eq}

\maketitle

\section{Introduction}

\noindent
To produce new exotic nuclei, there are several methods, such as fission \cite{fis1,fis2},
fragmentation/spallation \cite{frag1,frag2,frag3,frag4,frag5}, and fusion-evaporation reactions
\cite{fusion1,fusion2,fusion3}, are widely used in different laboratories over the world. However,
these reactions are not possible to produce the new neutron-rich nuclei in the region of trans-uranium
and superheavy nuclei \cite{isodis}. For the region of trans-uranium and superheavy nuclei, the limitations
of fusion-evaporation reactions have led to an exploration of multi-nucleon transfer (MNT) reactions
between projectile and target at energies around the Coulomb barrier
\cite{adt1,adt2,adt3,adt4,adt5,adt6,adt7,adt8,adt9,adt10,adt11,adt12}.

MNT reactions happen in quasi-elastic scattering process,
deep inelastic scattering, and partly quasifission reactions.\ The mechanisms of these reactions
are totally different.\ Deep inelastic scattering was discovered in 1960s. This new mechanism,
accompanied with the discovery of quasifission mechanism, greatly promoted the development of
heavy ion nuclear reactions.\ The 70s and 80s of the last century is  one of the rapid development period
of heavy ion nuclear physics \cite{linbook,bass80}. The interest of multinucleon transfer process
existed in quasi-elastic scattering process was renewed in later of 1980s \cite{von87,oertzen92}.
It was proposed that cold multi-proton transfer can be observed if appropriate dynamical
conditions are chosen, which will be helpful for the production of very heavy elements.
The mechanisms of deep inelastic transfer and quasi-elastic scattering process are quite different.
In deep inelastic transfer reactions, the transfer is mainly result of the friction mechanism,
accompanied by large kinematic energy dissipation. The total kinetic energy loss (TKEL)
can be hundred MeV, and the angular distribution covers a wide range from positive angle
to negative angle, which is called as hot multinucleon transfer
reaction.\ The energy dissipation of MNT reaction in quasi-elastic transfer reaction is small,
and the angular distribution is concentrated near the grazing angle,
which is called as cold multinucleon transfer reaction. For specific reactions, the distinguish
of these two kinds of mechanisms is unrealistic. The role of quasifission in multinucleon
transfer reaction was indicated in some experimental studies \cite{wolfs87,corradi99}.
It is also difficult to distinguish clearly deep inelastic scattering fragments with the
quasifission fragments \cite{bass80,corradi}.

\section{Experimental progress}

\noindent
The discovery of deep inelastic scattering can be traced back to the the pioneering work
performed around 1960 at the Yale University Heavy-Ion Linear Accelerator by Kaufmann and Wolfgang
\cite{kaufmann59}. Projectiles including $^{12}$C, $^{14}$N, $^{16}$O, $^{19}$F, are used to bombard
Al, Cu, Sn, Rh and other targets from near of Coulomb barrier energy to 10 MeV/u.\ Unexpected
phenomena were observed which were not consistent with the previously familiar patterns of compound or
direct reactions.\ Except that angular distribution of the single neutron transfer product
$^{15}$O is peaked  near the grazing angles, the cross section for (pn), (p2n) and (2p3n)
transfers are appreciable compared to the single nucleon transfers. The angular distributions
of these products are all peaked  at forward angles.\ These experiments indicated that
there may be a new reaction mechanism.\ At that time, they proposed a grazing contact
mechanism.\ It is considered that the bomb nucleus penetrates the target surface without
fusion, but rotated along the surface.\ The contact time is less than
half of the rotation period. Nuclear collective excitation and nucleon
transfer will happen under the strong friction of this process.\ Many instructiveness
conceptions are proposed in this model, such as the intermediate complex, negative
angle rotation, neck formation and elongation, surface tension, friction mechanisms
and so on. These conceptions laid the theoretical basis for deep inelastic scattering,
which has evolved into the friction model \cite{hasse78,gross78}, diffusion model and
transport model \cite{norenberg74}, etc.

In the 1960s, due to the limitations of accelerators, people could only use light heavy
ion beams (mainly C, N, O).\ The incident energy for these experiments were conducted
at generally 6--10 MeV/u.\ Many different experimental groups had observed similar multinucleon
transfer and forward angle  concentration  phenomenon \cite{galin70}.\ These experimental studies were
aimed both at a better understanding of the transfer reaction mechanism and at the production
of new isotopes.\ With the development of accelerators  at Dubna, Orsay, and Berkeley
in 1970s, heavier ions such as  Ar, Kr, Xe, etc., can be accelerated to higher energy.\ In addition,
the development of more advanced experimental techniques  were achieved, especially related
to correlation measurement such as angle-energy correlation, mass-energy correlation and so
on.\ For example, the development of the magnetic spectrometer in front of the particle telescopes
significantly improved the isotopic resolution by Artukh {\it et al}.\ \cite{artukh69}.\ The
substantial progress for the experimental studies eventually lead
to the establishment of this new reaction mechanism.

One of the most representative examples of deep inelastic scattering is the $^{40}$Ar+$^{232}$Th
experiments performed at Dubna \cite{artukh73}. Lately, these experimental results are perfectly
interpreted by Wilczy{\'n}ski on the basis of friction theory and scattering into negative angles
\cite{wilczynski73}.\ The contour plots of double differential cross section as a function of
centre of mass angle and total kinetic energy gives quite instructive explanation on this
phenomenon, which is often referred to as ``Wilczy{\'n}ski  plot''. The angular distributions of
the classical $^{40}$Ar+$^{232}$Th reactions shows an obvious peak around the grazing angle
for the fragments close to the projectile.\ The fragments far from the projectile contributes a
smooth part of the angular distribution.\ The bump and smooth part of the angular distribution
corresponds to the high energy (quasi-elastic) and low-energy (deep inelastic)
parts, respectively.\ There are two distinguished ridges in the double differential
cross sections contour plot with respect to the energy and angle,
which tends to cross near zero degrees. It was reasonably explained by
Wilczy{\'n}ski that there are results of the attractive nuclear forces
which renders the grazing trajectory towards smaller angles and then
to negative angles in deep inelastic process.\ An important gross feature of deep inelastic
transfer reactions is that a broad distribution of product masses can be observed. This is clearly shown
by the contour plot of double differential cross section with respect to mass and energy distributions
firstly introduced with kinematic coincidence technique \cite{tamain75}.

Based on the properties of deep inelastic transfer reaction, it was soon be used to generate new
isotopes during the following decades. Between 1969 and 2009, 77 new
isotopes had been found by using deep inelastic reactions or multinucleon
transfer reactions  by different particle identification methods at Dubna, Orsay, Berkeley, and Legnaro
\cite{isodis}.\ For light particles, it is relatively easy to be identified directly.
47 new neutron-rich light isotopes were discovered by using magnetic spectrometers and direct
particle  identification method.\ Between 1969 and 1971, 28 new neutron rich
isotopes were detected by Artukh {\it et al}., using the U300 heavy-ion cyclotron and
magnetic spectrometer combined with a $\Delta E$--$E$ semiconductor  telescope including
two semiconductor detectors at Dubna \cite{artukh69}.\ Auger and Guerreau {\it et al}.\ identified
more than ten heavier neutron-rich isotopes in deep inelastic reactions
induced by $^{40}$Ar on $^{238}$U with different incident energy, by measuring the
time-of-flight of the fragments and using two $\Delta E$ measurements with the
ALICE facility at Orsay around 1980 \cite{auger79,guerrau80}.
Similar experiment was performed with the Super HILAC accelerator at Berkeley
months later, with deep inelastic reactions $^{56}$Fe+$^{238}$U, and thirteen new
isotopes were discovered \cite{breuer80}.\ In Legnaro Tandem XTU-ALPI accelerator,
deep transfer reaction $^{82}$Se+$^{170}$Er was performed and $^{170}$Dy was deduced from
a 163 keV $\gamma$-ray spectra \cite{soderstrom10}.

The directly identification of charge and mass of very heavy fragments is
very difficult. They cannot be identified by alpha decay tagging because of the spontaneous
fission.\ It is also impossible to directly measure the mass and charge by measurement
of energy loss and time-of flight due to the  plasma effects and the resulting
pulse height deficit at Coulomb barrier region \cite{heinz18}.\ The missing mass
method was firstly introduced to identify the target-like fragments in multi-nucleon
transfers. The mass of target-like fragment can be deduced from the two-body kinematic
conditions of the projectile-like fragment. Chemical separation approach was also used to
identify target-like fragments. Based on these indirect methods, 9 target-like isotopes were
determined in multi-nucleon transfer reactions from 1950s to 1990s. As the earliest discovered
isotope in multi-nucleon transfer reaction, $^{13}$B was determined by measuring
protons with a CsI(Tl) scintillating crystal at the Van de Graaff accelerator of
the Enrico Fermi Institute for Nuclear Studies \cite{allison56}. Isotope $^{68}$Ni
was detected in 1977 with the Q3D spectrometer at the MP tandem of the
Max Planck Institute f{\"u}r Kernphysik \cite{bhatia77},
and $^{69}$Ni was detected later with with the double-focusing magnetic spectrometer
(BACCHUS) at the MP tandem of Orsay \cite{dessagne84}.\ Zhao {\it et al}.\ discovered a new nucleus $^{199}$Ir
using the transfer reaction $^{198}$Pt+$^{18}$O at 140 MeV with the high-resolution QMG/2
magnetic spectrometer at the Nuclear Structure Facility of Daresbury Laboratory \cite{zhao93}.
At the Heavy Ion Research Facility of the Institute of Modern Physics in Lanzhou (HIRFL),
five more isotopes of heavy elements, including $^{208}$Hg, $^{239}$Pa, $^{209}$Hg, $^{186}$Hf,
$^{238}$Th, were measured firstly in multinucleon transfer reactions with
relatively higher incident energy between 1993 and 1998 by chemical separation method
\cite{zhang94,yuan95,zhang98,yuan98,he99}.

The isotope separation on-line (ISOL) method, original used for identifying fragments for
fission and spallation reactions, was also used for distinguishing new isotopes
produced in multinucleon transfer reactions. The GSI ISOL facility was
operated  at the heavy-ion accelerator UNILAC at Darmstadt since 1976
and greatly improved afterwards \cite{bruske81}.\ During 1980s, 18 new neutron-rich
isotopes were identified with different heavy projectiles bombarding on
tungsten Ta/W targets or targets with tungsten-tantalum layers \cite{isodis}. The online
mass separator was used to distinguish these new isotopes. The heaviest nucleus
among these new isotopes are $^{232}$Ac and $^{234}$Ac, which are produced with
$^{238}$U projectile at incident energy 11.4 MeV/u \cite{gippert86}.\ Around 1990,
3 new isotopes were discovered in deep inelastic reactions by using the ISOL method
with the Super HILAC and the OASIS mass separation facility  at Berkeley
\cite{chasteler89,chasteler90}.

The cold  multi-nucleon transfer and multi-pair transfer were firstly
introduced by Oertzen.\ Based on experimental and theoretical hints,
several suggestions were formulated in order to observe the transfer of more than
four protons on a $^{248}$Cm target based on multiple proton-pair transfer
\cite{von87,oertzen92}.\ Transfer of multiple pairs will provide
critical information on nucleon-nucleon correlations, especially if
measurements are performed below the Coulomb barrier.\ A series of experiments
performed by Corradi {\it et al}.\ at the Tandem+ALPI accelerator complex of the
Laboratori Nazionali di Legnaro provided lots of evidence about the
pair or cluster transfer in multinucleon transfer reaction \cite{corradi}.
Reaction $^{40}$Ca+$^{124}$Sn at incident energy 170 MeV around Coulomb barrier was
measured with the newly developed time-of-flight magnetic spectrometer at Legnaro
in 1996 \cite{corradi96}.\ The TOF spectrometer was equipped with two microchannel
plate detectors and a multiparametric ionization chamber of $\Delta E$--$E$ type,
which were used for nuclear charge and energy measurement.\ A rotating scattering
chamber is also connected for measuring the angular distributions at six angles
near the grazing one.\ Angular and $Q$-value distributions for
a large variety of projectile-like charge and mass partitions produced
in the reaction were measured clearly.\ After comparison with independent
single-nucleon transfer modes, it was shown that more complex mechanisms shall be considered
for the larger drift towards neutron stripping.

With the same experimental setup at Legnaro, multinucleon transfer reactions
experiments $^{48}$Ca+$^{124}$Sn was performed to clarify the complex mechanisms
in 1997 \cite{corradi97}.\ It was demonstrated that, taking into account only
independent single-nucleon transfer modes, experimental isotope cross sections
for the $-2$p  and $+2$p channels cannot be described well.\ The discrepancies
can be caused by nucleon pair and cluster degrees of freedom.\ The MNT detection
equipment had been improved a lot during last two decades.\ The last generation large
solid angle magnetic spectrometers PRISMA \cite{stefanini02}, VAMOS \cite{savajols99}
and MAGNEX \cite{cunsolo02} greatly improved the detection efficiency more than an
order of magnitude than previous works. These spectrometers, companied with large
gamma arrays allows to detect directly and uniquely gamma-particle and nuclei
far from stability, produced via nucleon transfer or deep-inelastic reactions
especially in the neutron-rich region \cite{corradi13}.\ Many multinucleon transfer
reactions have given experimental hints on the pair transfer based on above mentioned
advanced experimental equipments, such as $^{64}$Ni+$^{238}$U \cite{corradi99},
$^{58}$Ni+$^{208}$Pb \cite{corradi02}, $^{40}$Ca+$^{208}$Pb \cite{szilner05},
$^{64}$Ni+$^{116}$Sn \cite{montanari14}, as well as the reaction
$^{16}$O+$^{208}$Pb performed at ANU \cite{evers11}.

Due to the available of high-quality $\gamma\gamma$-coincidence data
with anti-Compton-shielded multidetector arrays, the resolving
of fragments in deep inelastic reactions had become possible in 1990s
\cite{broda94}.\ Corresponding experimental spectroscopic studies on the
emitted fragments improved the study of nuclear structure as a function of
isospin and interesting phenomena like shell structure and magic number evolution
for neutron-rich nuclei.\ The gamma spectroscopic studies also inspired
the study of $N/Z$ equilibration process in the damped heavy-ion transfer
reaction \cite{broda06}.\ Based on different kinds of theoretical predictions
and experimental hints \cite{dasso94,adamian05,broda06,zagrebaev08-OYGxl,feng09-u,adamian10,adamian_10},
multinucleon transfer reaction was predicted to be able to produce
neutron-rich isotopes of heavy nuclei.\ The studies on this aspect have aroused
widely concern.\ And it seems that MNT reaction is the only current
available approach to access the region of  heavy  neutron-rich nucleus
\cite{corradi13}.\ It is also demonstrated that MNT reaction is a powerful method
promising to synthesize neutron rich nuclei around $N=126$, which are relevant for
the rapid neutron capture process and important for the understanding
the synthesize of the nuclei heavier than iron  in nature
\cite{zagrebaev08-OYGxl,watanabe_15}.\ Many MNT experiments have been carried on
in recent ten years aimed at production of new isotopes \cite{comas13,kozulin12,kratz13}.

In Ref.~\cite{zagrebaev08-OYGxl}, it is proposed that for production of possible last
``waiting point'' heavy neutron-rich nuclei located along the neutron closed shell $N=126$,
multinucleon transfer reactions in low-energy collisions of $^{136}$Xe+$^{208}$Pb shall be
explored.\ This reactions system at different incident energies had been studied at Dubna in 2012
\cite{kozulin12} and Argonne National Laboratory in 2015 \cite{barrett15},
respectively.\ Many isotopes had been found in both experiments,
especially over 200 projectile-like fragments (PLFs) and target-like
fragments (TLFs) were yielded from the these laboratories.\ These experiments
provide great opportunities for the testify of different theoretical models.
Another $^{136}$Xe+$^{208}$Pb  MNT reaction performed at the Laboratori Nazionali
di Legnaro proves the existence of a long-lived isomer in $^{133}$Xe \cite{vogt17}.\ Multinucleon
transfer reactions $^{156,160}$Gd+$^{186}$W were also studied recently to investigate $N=126$
shells \cite{kozulin17}.\ However, no new neutron-rich heavy nuclei was discovered yet.

By virtue of experimental multinucleon transfer reaction $^{136}$Xe+$^{198}$Pt
at 7.98 MeV/u, experimental verification was found firstly, which proves the theoretical
predictions that MNT would be the suitable tool to populate and characterize neutron-rich
isotopes around $N=126$ shell closure \cite{watanabe_15}.\ The experiment was carried out at
at GANIL using the large acceptance VAMOS++ spectrometer, and the EXOGAM array with ten
CLOVER germanium detectors surrounding the target.\ In this experiment, the first direct
measurements of the absolute production cross sections for a large number of
fragments were carried out.\ There were experimental hints for the production of $N=126$
nuclides $^{200}$W and $^{201}$Re in this work.\ However, these nuclides are deduced from
energies and angles of the projectile-like fragments, and the authors did not
claim the discovery of these isotopes \cite{isodis}.\ The work
demonstrated the possibility of producing the new neutron-rich isotopes around
$N=126$ shell closure by MNT reactions at both present and
future facilities.\ After comparison of the production rates in
MNT reactions with that from projectile fragmentation reaction
$^{208}$Pb+Be at 1 GeV/nucleon, it was indicated that the former one is
a more optimum way in the production of neutron-rich isotopes
around $N=126$ shell closure.\ However, by universally
comparison of the theoretical expected yields, beam intensities,
target thickness and detection efficiencies for Au isotopes produced
by multinuclear transfer reaction $^{64}$Ni+$^{207,208}$Pt and fragmentation
methods in Ref.~\cite{heinz18}, it is concluded that
fragmentation reactions appear much more favorable for
the production of neutron-rich nuclei along the $N=126$ shell.
Finally, it can be concluded that systematic studies to find
the optimum projectile-target combinations are critical
for the production of nuclei at this region.

For the production of new neutron-rich superheavy nuclei,
fragmentation reactions are not applicable, because there are no target or
projectile heavy enough.\ Fusion reactions is also not possible due to that
only neutron-deficient heavy nuclei can be produced by this way.\ Hence, MNT
reaction is the only possible way to access these nuclei.\ It is also predicted
in theoretical papers that ``asymmetry-exit-channel quasifission reactions''
\cite{adamian05} or ``inverse quasi-fission'' \cite{zagrebaev06} will be
potential methods to produce neutron-rich superheavy nuclei.\ Discussions on
the  prospects of this method from experimental view can be seen in
Refs.~\cite{corradi13,kratz15,schadel16,heinz18}.\ Many experiments have been
performed in recent years, such as $^{238}$U+$^{238}$U and $^{238}$U+$^{248}$Cm
carried out at the UNILAC accelerator of the GSI \cite{kratz13},
$^{136}$Xe+$^{238}$U performed at the INFN Laboratori Nazionali di Legnaro
\cite{vogt_15}.\ Important conclusions about the reaction dynamics
of these reaction systems have been drawn, but no sizable
yield of new neutron-rich heavy isotopes is found yet.\ One of the key bottlenecks
is that it is difficult to identify these high $Z$ elements by current experimental
detection methods, due to the Coulomb barrier plasma effects and
the resulting pulse height deficit problem \cite{heinz18}.\ New experimental
approaches shall be furthur developed, such as selective laster ionization,
Penning trap or multiple reflection time-of-flight mass spectrometer,
E-TOP telescopes with calorimeter.

One of the breakthrough in producing new isotopes by multinucleon transfer reaction
is the discovery of new neutron-deficient isotopes in 2015.\ Five new neutron-deficient
isotopes, including $^{216}$U, $^{219}$Np, $^{223}$Am, $^{229}$Am, and $^{233}$Bk were
discovered in the MNT reaction $^{48}$Ca+$^{248}$Cm at GSI
\cite{devaraja15}.\ The same reaction was performed  at GSI in 1986,
but no new isotopes were found at that time \cite{gaggeler86}.\ The new experiment
was performed by using the velocity filter SHIP, and the isotope
was identified via the $\alpha$ decay chains.\ By comparison of the product
cross section and experimental efficiency \cite{heinz18}, which are experimentally
relevant parameter and reflect the event count rate, it is shown that MNT reactions
are currently very promising ways for the production of new neutron deficient
transuranium isotopes due to their broad excitation functions.

\section{Theoretical progress}

\subsection{Dinuclear system (DNS) model}

\noindent
In the DNS model, after capture happens, nucleon transfer is affected by the relative motion
and the transfer process is considered as time-dependent.\ The evolution of the DNS is a diffusion
process along the mass asymmetry degree and simultaneously in the relative distance between the
centers of the interacting nuclei. The distribution probability can be obtained by solving a set
of master equations numerically in the potential energy surface \cite{DNS1,DNS2,DNS3,ref4}.

The total production cross section of a primary fragment with charge $Z$ and mass
number $A$ can be calculated as follows:
\begin{eqnarray}
\label{dc}
\sigma_{tr}^{DNS}(Z,A,E)&=&\frac{\pi\hbar^2}{2\mu E}\sum_l(2l+1)\nonumber\\
&&\times P_{c}(E,l)P^{DNS}_{tr}(Z,A,E),
\end{eqnarray}
where $P_{c}$ is capture probability and $P^{DNS}_{tr}(Z,A,E)$ is the production
probability of a fragment,\ $\mu$ is the reduced mass of the colliding system.

Nowadays, the MNT reactions have been proposed to synthesize the exotic nuclei in
neutron-rich heavy nuclei, neutron-rich superheavy nuclei, and neutron-deficient heavy
nuclei regions.\ The production cross sections of these exotic nuclei have been extensively
predicted by the DNS model. Usually, the neutron-rich heavy exotic nuclei can be produced
in the collisions of a neutron-rich stable ($^{64}$Ni, $^{136}$Xe, and $^{176}$Yb, etc.)
or radioactive ($^{132}$Sn, $^{139}$Xe, and $^{144}$Xe, etc) projectile with very heavy
target ($^{198}$Pt, $^{208}$Pb, and $^{238}$U, etc.). In Fig.~1, the calculated yields of
heavy neutron-rich nuclei in the transfer reaction $^{176}$Yb+$^{238}$U are shown
\cite{ref1}.\ The de-excitation process of primary fragments are treated with the GEMINI
code \cite{gemini}.\ Figure 1(a) shows the isotopic distributions at different incident
energies.\ The height of the interaction potential at the touching configuration is about
565 MeV.\ It is found that cross sections are relatively larger in neutron-rich side when
$E_{\textrm{c.m.}}=600$ MeV.\ The calculated cross sections for production of Eu, Tb, Ho,
and Yb isotopes in $^{176}$Yb+$^{238}$U reaction at $E_{\textrm{c.m.}}=600$ MeV
are shown in Fig.~1(b). The calculated production cross sections for the neutron-rich nuclei
$^{165,166,167,168}$Eu are 2.84 $\mu$b, 0.78 $\mu$b, 19.64 nb, and 1.36 nb; $^{169,170,171,172,173}$Tb
are 6.90 $\mu$b, 3.70 $\mu$b, 0.12 $\mu$b, 44.44 nb, and 1.00 nb; $^{173,174,175,176,177,178}$Ho are
46.24 $\mu$b, 16.83 $\mu$b, 2.00 $\mu$b, 0.85 $\mu$b, 73.75 nb, and 18.08 nb;
and $^{181,182,183,184,185}$Yb are 53.61 $\mu$b, 4.67 $\mu$b, 2.85 $\mu$b, 84.70 nb, and 4.29 nb,
respectively.\ It can be found that the cross sections of the $^{165}$Eu, $^{169}$Tb, $^{173}$Ho,
and $^{181}$Yb are quite large for experimental detection.

\begin{figure}[t!]
\centering
\includegraphics[width=3.5in]{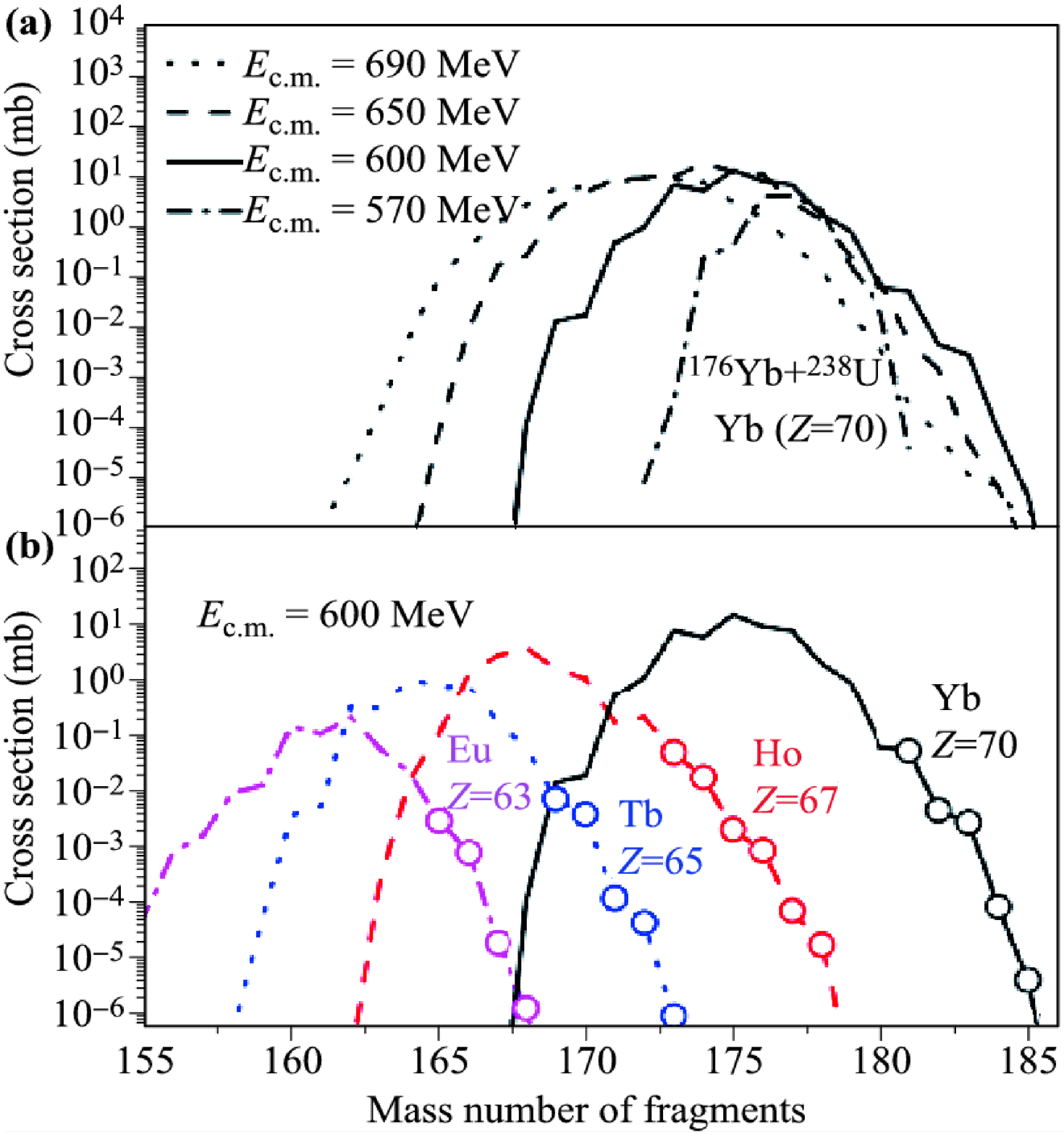}
\caption{{\bf (a)} Production cross sections of isotopes of Yb in the transfer reaction
$^{176}$Yb+$^{238}$U at $E_{\rm c.m.}=570$, 600, 650, and 690 MeV.\ {\bf (b)} Cross sections for
the formation of Eu, Tb, Ho, and Yb isotopes in the reaction $^{176}$Yb+$^{238}$U at
$E_{\rm c.m.}=600$ MeV.\ The circles denote the unknown neutron-rich nuclei.\ Reproduced
from Ref.~\cite{ref1}.}
\label{aba:fig1}
\end{figure}

The neutron-rich radioactive projectile has a larger neutron excess which could
significantly improve the production cross section of heavy neutron-rich nuclei.\ Figure 2
shows the calculated cross sections for reaction products with $Z=72$--$77$ in the collisions of
$^{136}$Xe, $^{139}$Xe, $^{144}$Xe, and $^{132}$Sn projectiles with $^{208}$Pb target \cite{ref2}.\ The collision
energies were adjusted in such a way that for all cases they are 1.1 times the corresponding potential
energies of contact (tip-to-tip) configurations of colliding nuclei.\ It can be seen that radioactive
beams enhance the production cross sections of neutron-rich nuclei. As can be seen the curves are
shifted to higher mass number region with increasing atomic number of surviving heavy nuclei.\ In our
calculation, it is noticed that many unknown neutron-rich nuclei can be produced with cross section of
$>$0.1 $\mu$b in radioactive induced transfer reactions and the production cross sections of unknown
neutron-rich nuclei in the reaction $^{144}$Xe+$^{208}$Pb are at least two orders of magnitude larger
than those in the reaction $^{136}$Xe+$^{208}$Pb.

\begin{figure}[t!]
\centering
\includegraphics[width=3.5in]{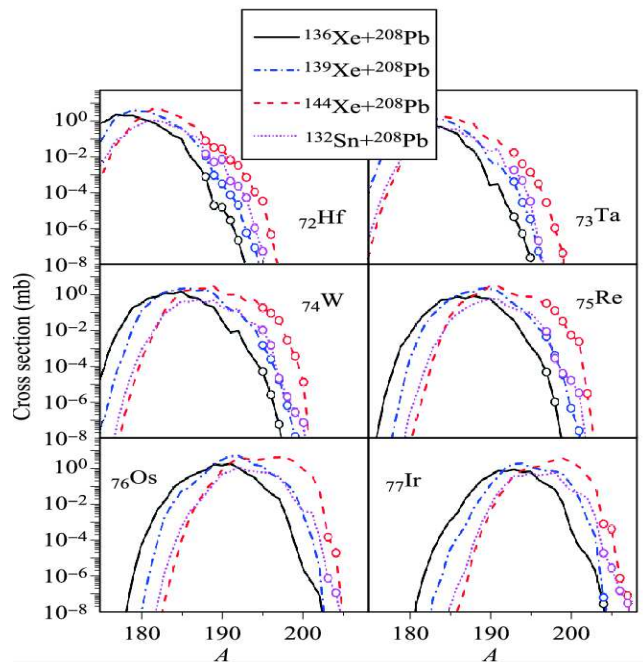}
\caption{Production cross sections of isotopes with $Z=72-77$ in the transfer
reactions $^{136}$Xe, $^{139}$Xe, $^{144}$Xe, and $^{132}$Sn projectiles with $^{208}$Pb.\ The incident
energies are chosen as 1.1 times the Coulomb barrier.\ The circles denote the unknown neutron-rich
nuclei.\ Reproduced from Ref.~\cite{ref2}.}
\label{aba:fig1}
\end{figure}

\begin{figure}[t!]
\centering
\includegraphics[width=3.5in]{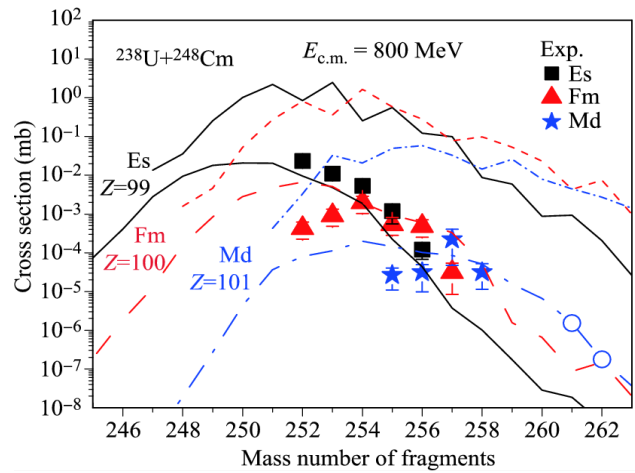}
\caption{Cross sections of Es, Fm and Md in $^{238}$U+$^{248}$Cm at $E_{\rm c.m.}=800$ MeV.\ The thin and thick
lines are distributions of primary and final fragments, respectively. The experimental data are taken from
Ref.~\cite{Cm}.\ The circles denote the unknown nuclei. Reproduced from Ref.~\cite{ref1}.}
\label{aba:fig1}
\end{figure}

\begin{figure}[t!]
\centering
\includegraphics[width=3.5in]{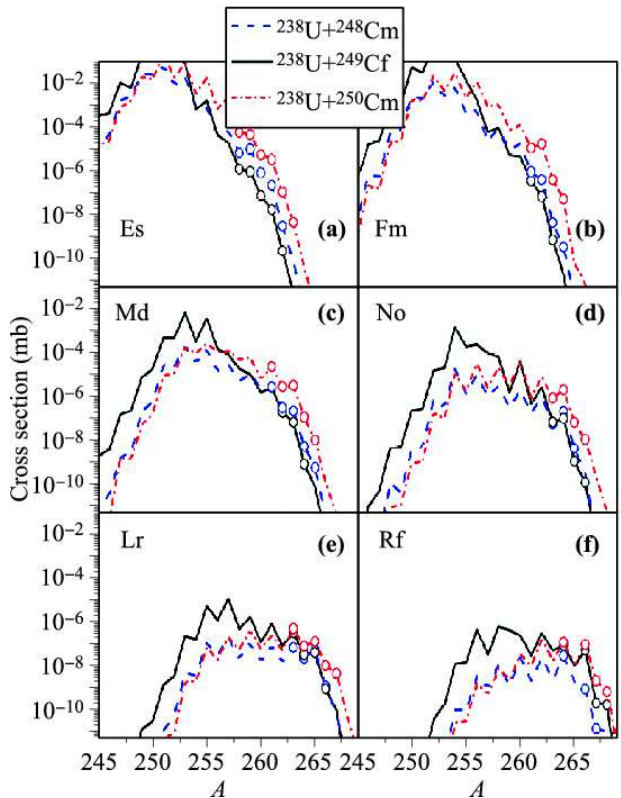}
\caption{Cross sections for synthesizing isotopes of the elements with $Z=99$--$104$ in transfer
reactions $^{238}$U+$^{248}$Cm, $^{238}$U+$^{249}$Cf, and $^{238}$U+$^{250}$Cm.\ The incident energy
$E_{\rm c.m.}=1.1 \times V_{CN}$. $V_{CN}$ is the interaction potential at the touching point (tip-tip).\ The circles
denote the unknown neutron-rich nuclei.\ Reproduced from Ref.~\cite{ref3}.}
\label{aba:fig1}
\end{figure}

The production of exotic nuclei towards the neutron-rich superheavy nuclei has been proposed
with damped collisions between two actinide nuclei.\ In Fig.~3, the production cross sections of Es
(+3p), Fm (+4p) and Md (+5p) in $^{238}$U+$^{248}$Cm at $E_{\rm c.m.}=800$ MeV are calculated by DNS model
\cite{ref1}.\ One can see a good agreement between theoretical and experimental for larger proton pickup
channels in approximately symmetrical system. The cross sections decrease drastically with increasing
atomic numbers of the fragments.\ The survival probabilities of most of the primary fragments are quite
low due to a dominant fission channel.\ The $^{261}$Md and $^{262}$Md could be synthesized in this reaction
with cross sections of about 1.52 nb and 0.17 nb, respectively.

In Fig.~4, we show the calculated cross sections for the production of the final reaction
products with $Z=99$--$104$ in damped collisions of $^{238}$U with $^{248}$Cm, $^{249}$Cf,
and $^{250}$Cm targets \cite{ref3}.\ The collision energies were adjusted in such a way that
for all three cases they are 1.1 times the corresponding potential energies of contact (tip-to-tip)
configurations of colliding nuclei. The $N/Z$ ratio of the target $^{248}$Cm is 1.58, which is higher
than 1.54 of $^{249}$Cf.\ The high $N/Z$ ratio target enhances the production cross sections of
transtarget neutron-rich nuclei.\ The reaction $^{238}$U+$^{248}$Cm shows much higher cross sections
than $^{238}$U+$^{249}$Cf for the production of unknown neutron-rich isotopes of the element Es.\ However,
for production of the transcalifornium nuclei, two more protons need to be transferred from $^{238}$U for
the reaction $^{238}$U+$^{248}$Cm in comparison with the reaction $^{238}$U+$^{249}$Cf. Thus,
it can be seen that the cross sections of unknown neutron-rich isotopes of the element Rf in the
reaction $^{238}$U+$^{249}$Cf become obviously higher than those in the reaction $^{238}$U+$^{248}$Cm.

\begin{figure}[b!]
\vspace*{-2mm}
\centering
\includegraphics[width=3.5in]{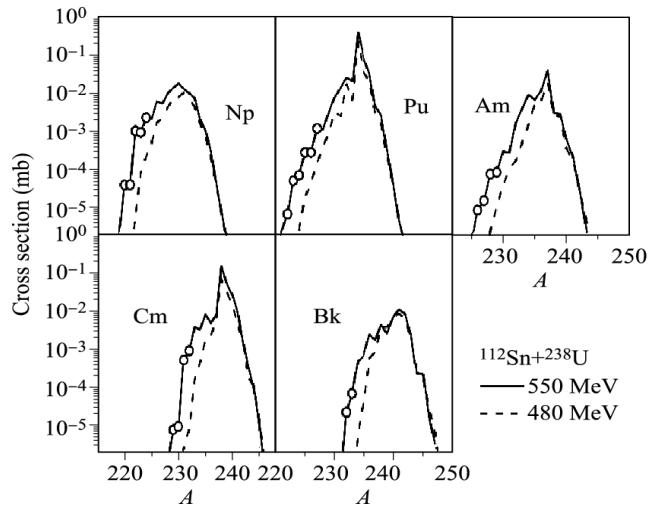}
\caption{Calculated cross sections for formation of actinide nuclei with $Z=93$--$97$ in the
reaction $^{112}$Sn+$^{238}$U with different incident energies. The circles denote the unknown
nuclei.\ Reproduced from Ref.~\cite{ref5}.}
\label{aba:fig1}
\end{figure}

To produce neutron-deficient actinide nuclei, a neutron-deficient projectile would be better.\ Figure 5 shows the
predicted cross sections for producing unknown neutron-deficient nuclei with $Z=93$--$97$ in the reaction
$^{112}$Sn+$^{238}$U \cite{ref5}. The yields of primary fragments are larger at higher incident energy.\ However,
with increasing excitation energies of primary fragments, more neutrons could be evaporated, which can counteract
the increased cross sections of primary fragments. The influence of incident energy in the neutron-deficient
region is pronounced. The unknown neutron-deficient nuclei $^{220}$Np, $^{221}$Np, $^{222}$Np, $^{223}$Np,
and $^{224}$Np can be produced in the reaction $^{112}$Sn+$^{238}$U at $E_{\textrm{c.m.}}=550$ MeV,
with cross sections 0.040, 0.041, 1.0, 0.96, and 2.3 $\mu$b, respectively. $^{223}$Pu, $^{224}$Pu,
$^{225}$Pu, $^{226}$Pu, and $^{227}$Pu can be produced with cross sections 0.053, 0.073, 0.30, 0.29, and
1.2 $\mu$b, respectively. The predicted cross sections of $^{228}$Am, $^{229}$Am, $^{231}$Cm, $^{232}$Cm,
$^{232}$Bk, and $^{233}$Bk are 0.079, 0.088, 0.52, 0.90, 0.022, and 0.066 $\mu$b, respectively.

\subsection{DNS+GRAZING model}

\noindent
GRAZING model is designed in order to evaluate the observable in the MNT
reactions that happen in the grazing region \cite{GRA1,GRA2,GRA3,GRA4}.\ In the GRAZING model,
exchange of particles is based on well-known form factors of the quantum coupled equations in a
mean field approximation, while the colliding nuclei are considered to move on classical
trajectories.\ Similar to the DNS model, the total production cross section can be expressed as
\begin{eqnarray}
\label{gc}
\sigma_{tr}^{GRAZING}(Z,A,E)=\frac{\pi\hbar^2}{2\mu E}\sum_l(2l+1)\quad\nonumber\\
\times(1-P_{c}(E,l))P^{GRAZING}_{tr}(Z,A,E).
\end{eqnarray}

It can be seen apparently from Eq.~(\ref{dc}) and Eq.~(\ref{gc}), the GRAZING model describes
transfer reactions without capture, which happen at the scattering process. While DNS model describes
the transfer reaction happens in the dinuclear evolution process after capture.\ These two models represents
two kind of transfer mechanisms, which are mutually complementary.\ Many experimental data indicate that MNT
reactions include the direct transfer reaction and DIT reaction.\ A schematic picture about the subdivision
of MNT reactions described by these two models is shown in Fig.~6.\ In this work, we try to simply
combine DNS and GRAZING model (i.e., $\sigma_{tr}^{tot}=\sigma_{tr}^{DNS}+\sigma_{tr}^{GRAZING}$)
for description of MNT reactions.

\begin{figure}[b!]
\vspace*{-2mm}
\centering
\includegraphics[width=3.5in]{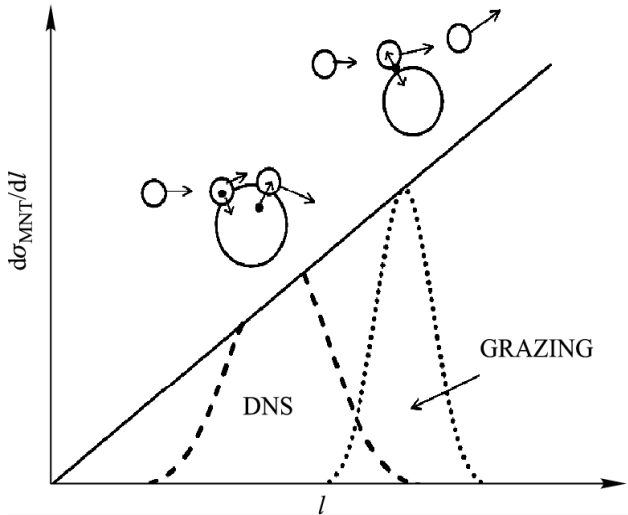}\\[-1mm]
\caption{Schematic picture of the distribution of the cross section differential with respect
to the angular momentum $l$ for MNT reactions by the GRAZING model and DNS model. Reproduced
from Ref.~\cite{ref6}.}
\label{aba:fig1}
\vspace*{-1mm}
\end{figure}

In order to contain simultaneously the scattering and capture process, we combine the DNS
and GRAZING model for description of MNT reactions.\ The isotopic production cross sections of $^{64}$Ni+$^{238}$U
at $E_{\rm c.m.}=307.5$ MeV are shown in Fig.~7 \cite{ref6}.\ For the larger proton stripping channels,
the DNS model agrees with experimental data better than GRAZING model. While for the $-1$p to $+1$p channels,
the GRAZING model is better than DNS model.\ It is hard to describe the transfer reaction with individual
scattering or capture mechanism. The GRAZING+DNS model makes a large improvement on $-6$p to $+1$p
channels.\ However, the cross sections of $+2$p channel are substantially underestimated.\ This combined model
should be further developed for describing multinucleon transfer reactions.

\begin{figure}[t!]
\centering
\includegraphics[width=3.5in]{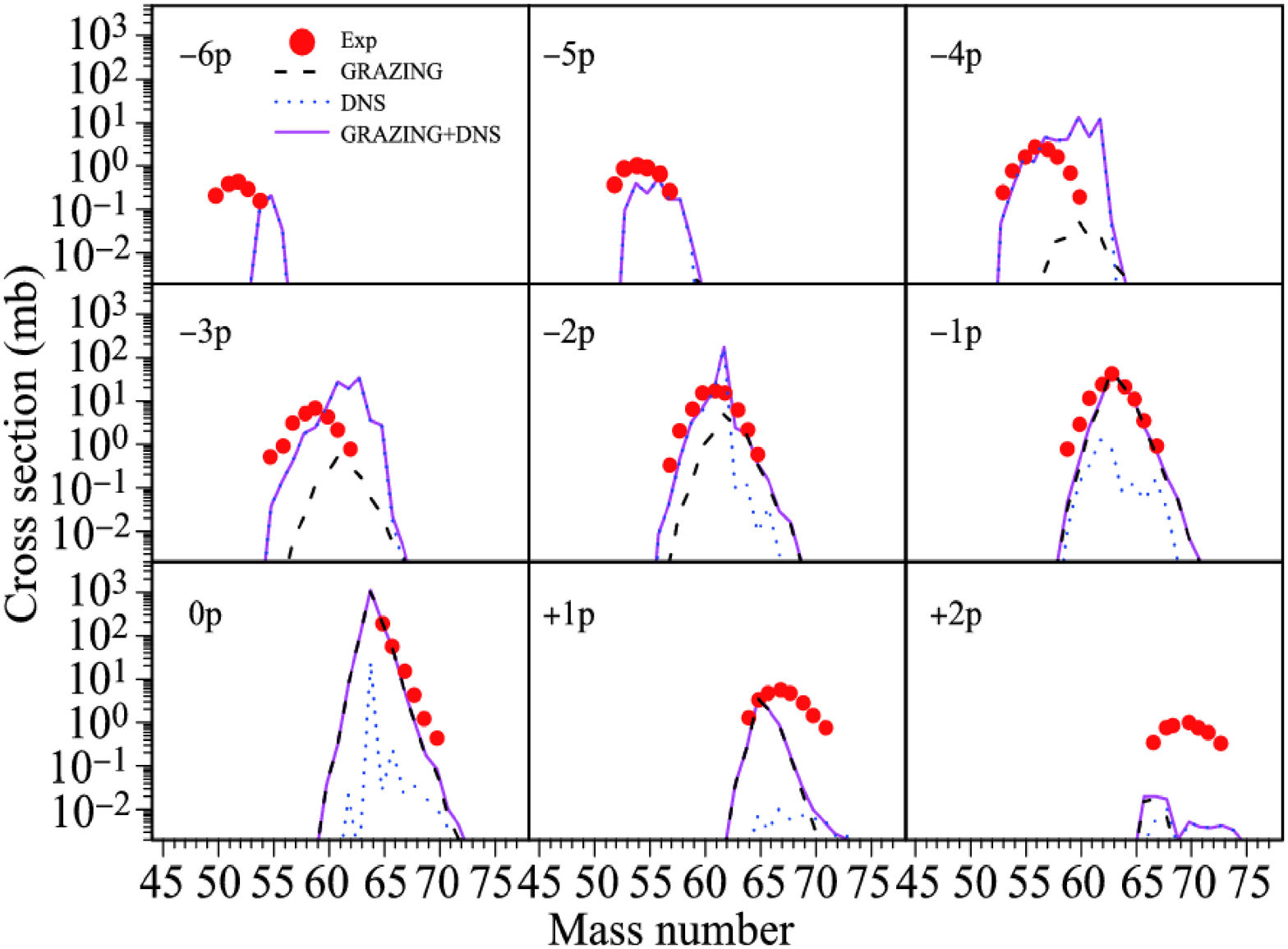}
\caption{Isotopic production cross sections (secondary fragments) of $^{64}$Ni+$^{238}$U at
$E_{\rm c.m.}=307.5$ MeV.\ The experimental data are from Ref.~\cite{corradi99}.\ Reproduced
from Ref.~\cite{ref6}.}
\label{aba:fig1}
\end{figure}

\subsection{ImQMD model}

\noindent
The ImQMD model is a semiclassical microscopic dynamics model which includes the mean-field and
nucleon-nucleon collisions as well as the Pauli principle \cite{QMD1}.\ In the ImQMD model, each nucleon
is represented by a coherent state, and the total $N$-body wave function is assumed to be the
direct product of coherent states. The standard Skyrme force with the omission of spin-orbit term is adopted
for describing not only the bulk properties but also the surface  properties of nuclei.\ Simultaneously,
the stochastic two-body collision process is added to the time evolution by the Hamilton equation of
motion.\ The final state of the two-body collision process is checked so that it obeys the
Pauli principle \cite{QMD2}.\ The time evolution of $\bm{r}_{i}$ and $\bm{p}_{i}$
for each nucleon is governed by Hamiltonian equations of motion
\begin{equation}
\dot{\bm{r}}_{i}=\frac{\partial H}{\partial \bm{p}_{i}},\quad
\dot{\bm{p}}_{i}=-\frac{\partial H}{\partial \bm{r}_{i}}.
\end{equation}
The Hamiltonian of the system includes the kinetic energy $T=\sum_{i}\frac{\bm{p}_{i}^{2}}{2m}$ and effective
interaction potential energy
\begin{equation}
H=T+U_{\rm Coul}+U_{\rm loc},
\end{equation}
where, $U_{\rm Coul}$ is the Coulomb energy, which is written as a sum of the direct and the exchange contribution
\begin{eqnarray}
U_{\textmd{Coul}}&=&\frac{1}{2}\int\!\!\!\int{\rho_{p}({r})}\frac{e^{2}}{|{r}-{r}'|}{\rho_{p}({r}')}{\rm d}{r}{\rm d}{r}'\\
&&-e^{2}\frac{3}{4}\left(\frac{3}{\pi}\right)^{1/3}\int\rho_{p}^{4/3}{\rm d}{r}.
\end{eqnarray}
$\rho_{p}$ is the density distribution of protons of the system.

The nuclear interaction potential energy $U_{\textrm{loc}}$ is obtained from the integration of the
Skyrme energy density functional $U=\int V_{\textrm{loc}}(\bm{r}){\rm d}\bm{r}$ without the spin-orbit term,
which reads
\begin{eqnarray}
V_{\rm loc}&=&\frac{\alpha}{2}\frac{\rho^{2}}{\rho_{0}}+\frac{\beta}{\gamma+1}
\frac{\rho^{\gamma+1}}{\rho {0}^{\gamma}}+\frac{\textsl{g}_{sur}}{2\rho_{0}}(\nabla\rho)^{2}\nonumber\\
&&+\frac{C_{s}}{2\rho _{0}}(\rho^{2}-\kappa _{s}(\nabla\rho)^{2})\delta^{2}+g_{\tau}\frac{\rho^{\eta+1}}{\rho_{0}^{\eta}}.
\label{12}
\end{eqnarray}
Here $\rho=\rho_{n}+\rho_{p}$ is the nucleons density.\ $\delta=(\rho_{n}-\rho_{p})/(\rho_{n}+\rho_{p})$
is the isospin asymmetry. The density distribution function $\rho$ of a system can be read
\begin{equation}
\rho(r)=\sum_{i}\frac{1}{(2\pi\sigma_r)^{3/2}}\exp\left[-\frac{(\bm r-\bm r_i)^2}{2{\sigma_r}^2}\right].
\label{aba:app1}
\end{equation}
The $\sigma_r$ is the wave-packet width of the nucleon in coordinate space.\ The ImQMD model is
successfully applied to heavy-ion fusion reactions, multinucleon transfer reactions and intermediate-energy
fragmentation reactions \cite{QMD2,QMD3,QMD4,QMD5}.\ More descriptions of the ImQMD model please see
Refs.~\cite{QMD1,QMD2}.

The isotopic production cross sections of $^{136}$Xe+$^{208}$Pb at $E_{\rm c.m.}=450$
MeV are shown in Fig.~8.\ One can see that GRAZING is a suitable model to estimate the
production cross sections only for $-1$p to $+2$p. It grossly underestimates the production
cross section by orders of magnitude in the case of more proton transfers.\ For the DNS model,
the height of peaks kept consistent with experimental data at least on the orders of magnitude
for $-3$p to $+2$p.\ For the ImQMD model, the height of peaks always kept consistent with experimental
data for $-3$p to $+3$p, but the peak widths are usually greater than experimental data especially for
$\Delta Z>0$.

\begin{figure}[t!]
\centering
\includegraphics[width=3.5in]{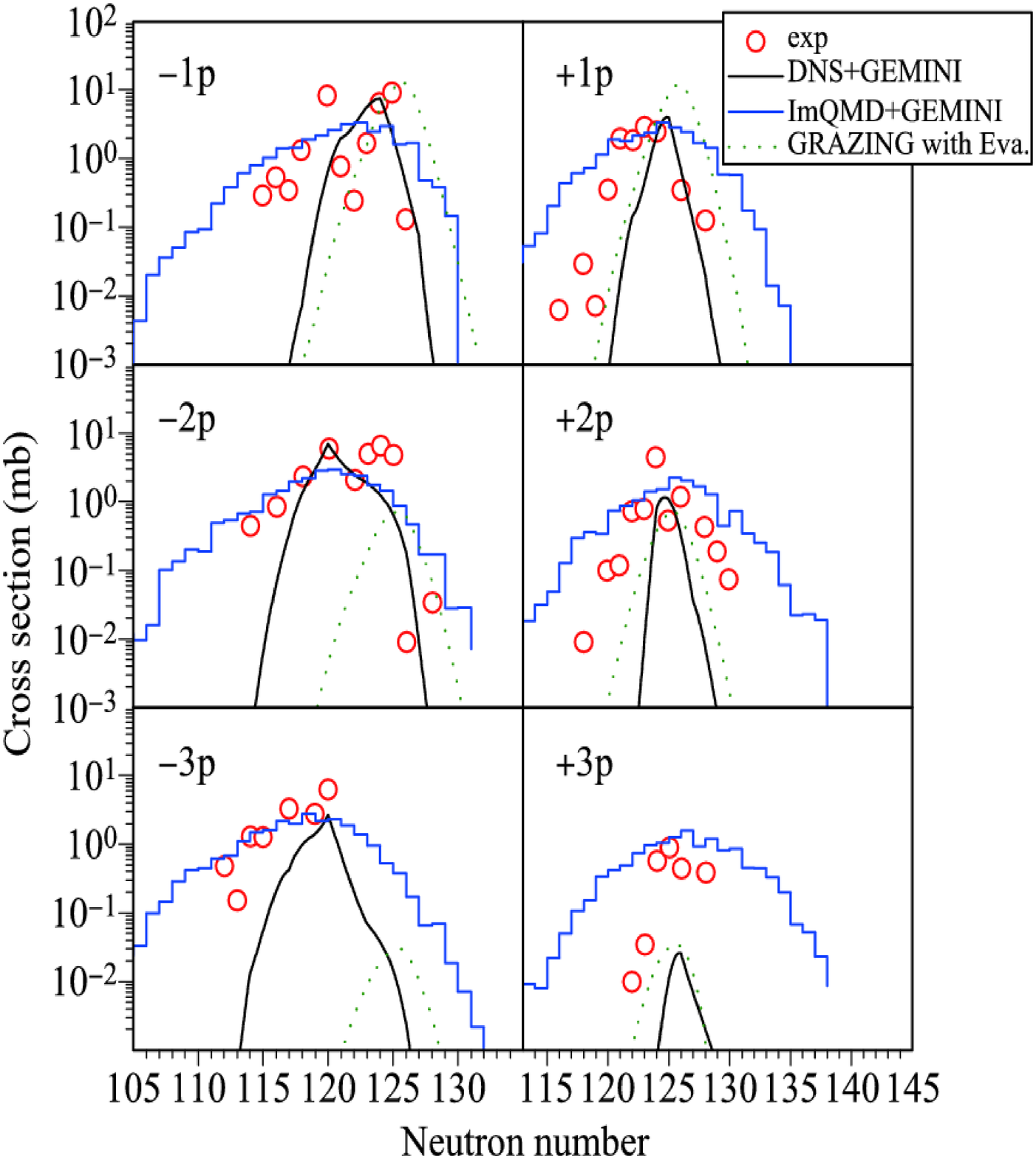}\\[-0.5mm]
\caption{Isotopic production cross sections (secondary fragments) of $^{136}$Xe+$^{208}$Pb.\ The experimental
data are from Ref.~\cite{barrett15}. Reproduced from Ref.~\cite{QMD4}.}
\label{aba:fig1}
\vspace*{-1mm}
\end{figure}

\begin{figure}[t!]
\centering
\includegraphics[width=3.5in]{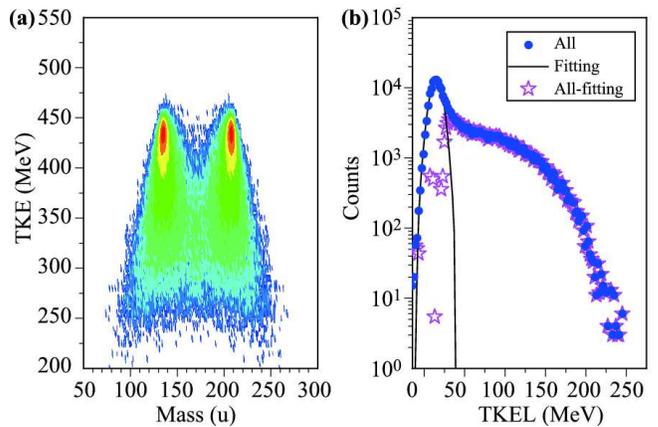}\\[-0.5mm]
\caption{Calculated TKE-mass distribution and TKEL distribution in $^{136}$Xe+$^{208}$Pb
reactions at $E_{\rm c.m.}=450$ MeV for primary binary fragments. In right panel, the solid circles
denote all the binary events; the solid line denotes the Gaussian fitting for the quasielastic events;
the stars are the difference between all the binary events and Gaussian distribution. Reproduced
from Ref.~\cite{QMD4}.}
\label{aba:fig1}
\vspace*{-1mm}
\end{figure}

In multinucleon transfer reactions, the nucleons transfer is accompanied by the energy
dissipation.\ The total-kinetic--energy--mass (TKE-mass) distribution can be used to judge
the reaction type.\ Figure 9(a) shows the TKE-mass distributions of $^{136}$Xe+$^{208}$Pb at
$E_{\rm c.m.}=450$ MeV for primary binary fragments \cite{QMD4}.\ The total kinetic energy (TKE)
is calculated in the center-of-mass system. The central collisions are deep-inelastic
reactions.\ There are a large numbers of nucleons transfers between the projectile and
target.\ The masses distribute in a rather broad range which is from 80 to 250.\ For
peripheral collisions, there are only a few nucleon transfers with smaller energy loss
because the reaction mechanism is dominated by quasielastic collisions.\ Figure 9(b) shows
that the distribution of the total kinetic energy lost (TKEL, i.e., $E_{\rm c.m.}$-TKE)
for primary binary events \cite{QMD4}.\ In Fig.~9(b) one can see that a Gaussian-type
distribution appeared at TKEL values lower than 40 MeV, which correspond to quasielastic
collision events.\ Most of the deep-inelastic events are located at TKEL values
above 40 MeV.\ An intense competition between the quasielastic and deep-inelastic reactions
caused by dynamical fluctuation can be find at the TKEL values near 40 MeV.\ The TKEL
distribution calculated by the ImQMD model is very similar to that measured by experiment
from reaction $^{88}$Sr+$^{176}$Yb at an incident energy slightly above the Bass barrier.

In Fig.~10(a), we show the average kinetic energy $\langle E_k\rangle$ of primary fragments for
$^{136}$Xe+$^{198}$Pt at $E_{\textmd{lab}}=7.98$ MeV/nucleon as a function of the impact parameters
\cite{QMD5}.\ For $12\leq b$ fm, the reaction mechanism is mainly elastic scattering.\ No energy is
lost in the system after the collisions. In the region of $8\leq b<12$ fm, the $\langle E_k\rangle$
value of the target-like fragments (TLFs) is around 150 MeV.\ While corresponding $\langle E_k\rangle$
value for the Projectile-like fragments (PLFs) decreases rapidly with decreasing impact parameters.\ The kinetic
energy of the projectile dissipate quickly to internal excitation energy in quasielastic and deep-inelastic
collisions.\ For $b<8$ fm, the $\langle E_k\rangle$ value for TLFs increases rapidly with the decreasing
impact parameters. A large amount of energy transfer from the projectile to target occurs in the
quasifission collisions.

\begin{figure}[t!]
\centering
\includegraphics[width=3.5in]{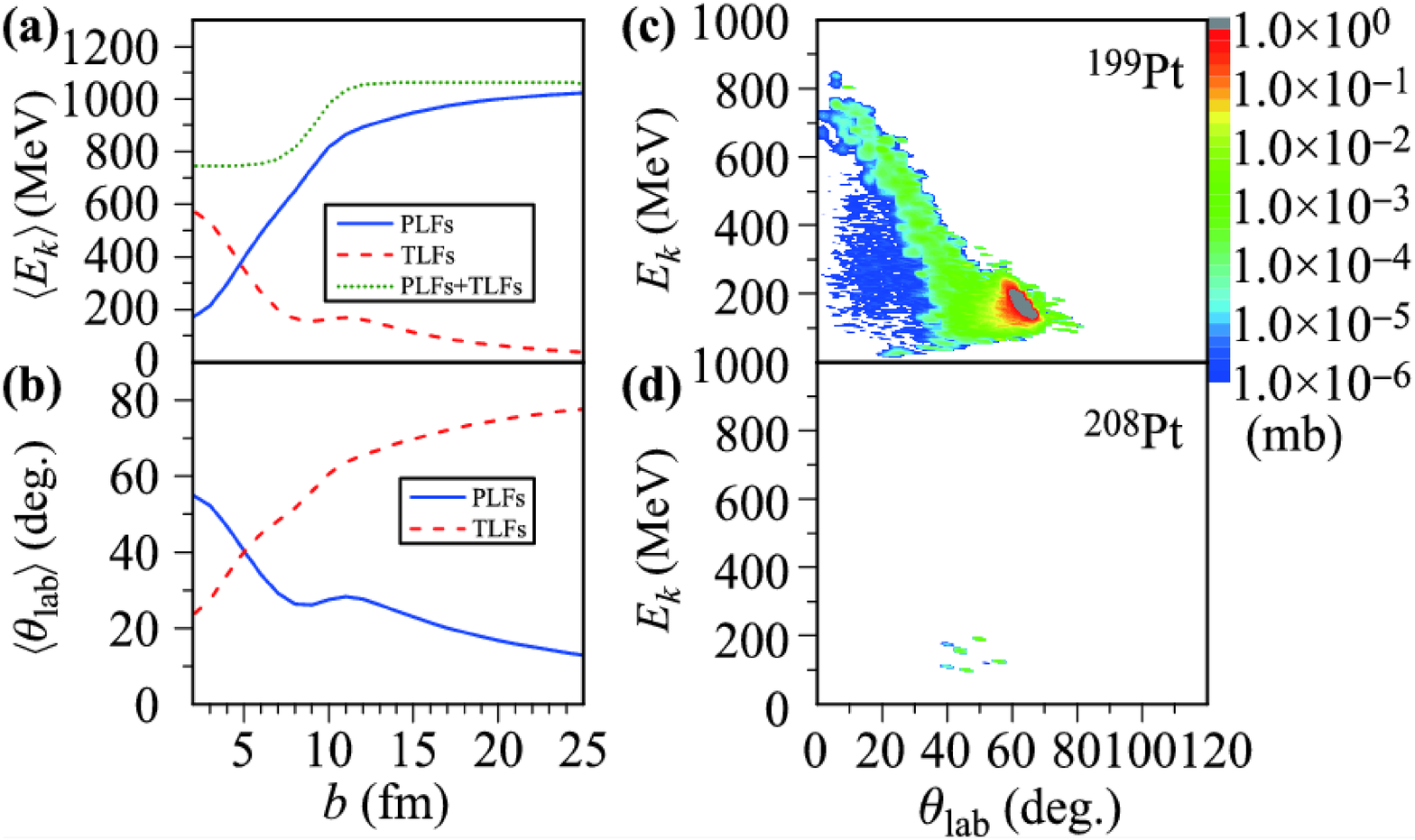}\\[-1mm]
\caption{Left panels: Average kinetic energy $\langle E_k\rangle$ and emission angle $\langle\theta_\textmd{lab}\rangle$
of primary fragments for $^{136}$Xe+$^{198}$Pt at $E_{\textmd{lab}}=7.98$ MeV/nucleon in the laboratory
frame as a function of impact parameters.\ Right panels: Calculated double differential cross sections
(secondary) of $^{199}$Pt and $^{208}$Pt. Reproduced from Ref.~\cite{QMD5}.}
\label{aba:fig1}
\vspace*{-1mm}
\end{figure}

The mean emission angles $\langle\theta_\textmd{lab}\rangle$ for the PLFs and TLFs at different
impact parameters for $^{136}$Xe+$^{198}$Pt at $E_{\textmd{lab}}=7.98$ MeV/nucleon are shown in
Fig.~10(b) \cite{QMD5}.\ The emission angle of light PLFs is significantly affected by nuclear
force and Coulomb force. The Coulomb force is dominant when impact parameter is greater than 12 fm. In the
region of $8\leq b<12$ fm, both the nuclear force and the Coulomb force are important to influence
the emission angle of light PLFs. This leads to the result that the values of $\langle\theta_{\textmd{lab}}\rangle$
for PLFs keep around at $28^\circ$ in quasielastic and in deep-inelastic collisions.\ For the TLFs,
the average emission angle decrease with decreasing impact parameters in the region of $2<b$ fm.

From Figs.~10(a) and (b), we can distinguish roughly the reaction mechanisms with impact
parameters and emission angles. The corresponding impact parameters for the elastic,
quasielastic, deep-elastic and quasifission collisions are in the region of $12<b$,
$10<b\leq 12$, $8<b\leq 10$, and $b\leq 8$ fm, respectively. The angles of TLFs in the
reaction mechanisms of elastic, quasielastic, deep-inelastic and quasifission distribute
in the region of $65^\circ$--$78^\circ$, $60^\circ$--$65^\circ$, $50^\circ$--$60^\circ$,
and $23^\circ$--$50^\circ$, respectively. Note that this partition for different reaction
mechanisms is rough.\ The border between different reaction mechanisms gets a bit fuzzy.\ However,
knowing the kinetic energy and the emission angle of a fragment, we can estimate
approximately its production mechanism.

Following the above analysis, we can investigate the production mechanism of the new neutron-rich
nuclei.\ The double differential cross sections of $^{199}$Pt and $^{208}$Pt are shown in
Figs.~10(c) and (d) \cite{QMD5}.\ The $^{199}$Pt (+1n) could be produced by three reaction
mechanisms.\ The greatest contribution to producing $^{199}$Pt is quasielastic collisions
(grey area).\ The corresponding emission angles are from $60^\circ$ to $70^\circ$ and
kinetic energy is around 150 MeV.\ The fragments with emission angles from $40^\circ$ to
$60^\circ$ and kinetic energy around 150 MeV is produced primarily in the deep-inelastic
collisions.\ The fragments with smaller emission angles and kinetic energy larger than 200 MeV
is mainly produced in quasifission mechanism.\ The very neutron-rich isotope $^{208}$Pt ($+10$n)
is produced in deep-inelastic reactions.\ The exotic neutron-rich nuclei cannot be generated in
quasielastic and quasifission collisions.\ Because in the quasielastic collisions, only a few
nucleons could be transferred.\ While for the quasifission collisions, large excitation energy
is obtained in energy dissipation processes which causes that the primary fragments evaporate
more neutrons in the deexcitation processes.

\begin{figure}[b!]
\vspace*{-2mm}
\centering
\includegraphics[width=3.5in]{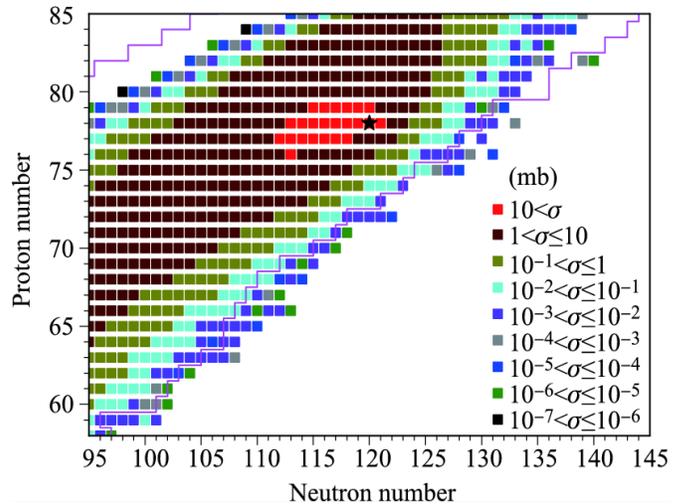}\\[-0.5mm]
\caption{Calculated production cross section distributions for the secondary fragments in the
$^{136}$Xe+$^{198}$Pt reaction. The folding lines denote the boundary of known nuclei.\ The star denotes
the position of the target. Reproduced from Ref.~\cite{QMD5}.}
\label{aba:fig1}
\vspace*{-1mm}
\end{figure}

Figure 11 shows that the final production cross section distributions for TLFs \cite{QMD5}.\ From Fig.~11,
one can see that about 50 new neutron-rich nuclei have survived after the deexcitation processes.\ The production
cross sections of these new neutron-rich nuclei are mainly located in the region of $10^{-3}$--$10^{-6}$ mb.\ Actually,
in the Ref.~\cite{watanabe_15}, these new neutron-rich nuclei were not detected because the detection limit of the
experiment is $10^{-2}$ mb.

\section{Predicated cross sections for new exotic\\nuclei}

\noindent
The predicted production cross sections of the new nuclei in MNT reactions with stable
projectile-target combination are listed in Table 1.\ The transfer reactions $^{136}$Xe+$^{186}$W
\cite{XeW}, $^{136}$Xe+$^{208}$Pb \cite{ref2}, $^{136}$Xe+$^{248}$Cm \cite{ref5}, and $^{112}$Sn+$^{238}$U \cite{ref5}
are calculated within the framework of the DNS model.\ The new isotopes in $^{136}$Xe+$^{198}$Pt \cite{QMD5}
reaction are predicted by using the ImQMD model.\ The $^{136}$Xe+$^{186}$W reaction can produce the new
neutron-rich nuclei with $Z=70$--$72$.\ The transfer reaction of $^{136}$Xe+$^{208}$Pb at $E_{\rm c.m.}=465$ MeV can
produce the new neutron-rich nuclei with atomic number form 72 to 75. The $^{136}$Xe+$^{248}$Cm reaction at
$E_{\rm c.m.}=510$ MeV can produce the new neutron-rich trans-uranium nuclei with $Z=92-96$.\ The neutron-deficient
$^{112}$Sn with $^{238}$U at $E_{\rm c.m.}=550$ MeV can produce the new neutron-deficient trans-uranium nuclei
with atomic number from 93 to 97. The reaction of $^{136}$Xe+$^{198}$Pt at $E_{\rm c.m.}=643$ MeV can produce
the new neutron-rich nuclei with $Z$ from 59 to 82.

\begin{table}[h!]
\doublerulesep 0.3pt
\tabcolsep 7.8mm
\setlength{\tabcolsep}{4pt}   %%%设置表格平均分配列间距
\renewcommand{\arraystretch}{1.25}  %%%设置表格行高
\caption{Predicted cross sections of new isotopes in MNT reactions from the DNS and ImQMD model.}
\vspace*{-1mm}
\footnotesize{\begin{tabular*}{86mm}{ccc}\hline
Reactions & Isotopes & Cross section $(\textrm{mb})$\\\hline\hline
\parbox[t]{42.5mm}{\centering $^{136}$Xe+$^{186}$W, $E_{\rm c.m.}=406$ MeV} & & DNS Cal.\\
 &  $^{181}$Yb   &  $6.659\times10^{-2}$ \\
 &  $^{182}$Yb   &  $2.135\times10^{-2}$ \\
 &  $^{183}$Yb   &  $2.030\times10^{-3}$ \\
 &  $^{184}$Yb   &  $1.813\times10^{-5}$ \\
 &  $^{185}$Yb   &  $7.659\times10^{-7}$ \\
 &  $^{186}$Yb   &  $4.265\times10^{-7}$ \\
 &  $^{185}$Lu   &  $2.160\times10^{-4}$ \\
 &  $^{186}$Lu   &  $4.037\times10^{-5}$ \\
 &  $^{187}$Lu   &  $1.289\times10^{-5}$ \\
 &  $^{188}$Lu   &  $2.270\times10^{-7}$ \\
 &  $^{188}$Hf   &  $1.481\times10^{-4}$ \\
 &  $^{189}$Hf   &  $1.028\times10^{-5}$ \\
 &  $^{190}$Hf   &  $3.061\times10^{-6}$ \\\hline
$^{136}$Xe+$^{208}$Pb, $E_{\rm c.m.}=465$ MeV & & DNS Cal.\\
 &  $^{188}$Hf   &  $6.958\times10^{-4}$ \\
 &  $^{189}$Hf   &  $1.710\times10^{-5}$ \\
 &  $^{190}$Hf   &  $1.368\times10^{-5}$ \\
 &  $^{191}$Hf   &  $2.543\times10^{-6}$ \\
\hline
\end{tabular*}
}
\renewcommand{\arraystretch}{1}  %%%设置表格行高
\end{table}

\newpage
\setcounter{table}{0}
\begin{table}[h!]
\doublerulesep 0.3pt
\tabcolsep 7.8mm
\setlength{\tabcolsep}{4pt}   %%%设置表格平均分配列间距
\renewcommand{\arraystretch}{1.29}  %%%设置表格行高
\caption{\hfill (continued)}
\vspace*{-1mm}
\footnotesize{\begin{tabular*}{86mm}{ccc}\hline
Reactions & Isotopes & Cross section $(\textrm{mb})$\\\hline\hline
 &  $^{192}$Hf   &  $2.032\times10^{-7}$ \\
 &  $^{193}$Ta   &  $4.048\times10^{-6}$ \\
 &  $^{194}$Ta   &  $4.668\times10^{-7}$ \\
 &  $^{195}$Ta   &  $2.200\times10^{-8}$ \\
 &  $^{195}$W    &  $4.739\times10^{-5}$ \\
 &  $^{196}$W    &  $2.472\times10^{-6}$ \\
 &  $^{197}$W    &  $5.820\times10^{-8}$ \\
 &  $^{197}$Re   &  $4.413\times10^{-5}$ \\
 &  $^{198}$Re   &  $1.003\times10^{-6}$ \\\hline
$^{136}$Xe+$^{248}$Cm, $E_{\rm c.m.}=510$ MeV & & DNS Cal.\\
 &  $^{244}$U    &  $3.543\times10^{-4}$ \\
 &  $^{245}$U    &  $4.610\times10^{-5}$ \\
 &  $^{246}$U    &  $1.423\times10^{-7}$ \\
 &  $^{247}$U    &  $2.659\times10^{-8}$ \\
 &  $^{248}$U    &  $2.032\times10^{-8}$ \\
 &  $^{245}$Np   &  $1.044\times10^{-3}$ \\
 &  $^{246}$Np   &  $1.706\times10^{-4}$ \\
 &  $^{247}$Np   &  $1.897\times10^{-6}$ \\
 &  $^{248}$Np   &  $8.901\times10^{-7}$ \\
 &  $^{249}$Np   &  $2.438\times10^{-7}$ \\
 &  $^{248}$Pu   &  $1.524\times10^{-4}$ \\
 &  $^{249}$Pu   &  $4.610\times10^{-5}$ \\
 &  $^{250}$Pu   &  $6.282\times10^{-6}$ \\
 &  $^{251}$Pu   &  $2.457\times10^{-6}$ \\
 &  $^{249}$Am   &  $3.362\times10^{-4}$ \\
 &  $^{250}$Am   &  $1.934\times10^{-4}$ \\
 &  $^{251}$Am   &  $1.387\times10^{-4}$ \\
 &  $^{252}$Am   &  $2.377\times10^{-5}$ \\
 &  $^{253}$Cm   &  $1.251\times10^{-4}$ \\
 &  $^{254}$Cm   &  $5.188\times10^{-5}$ \\
 &  $^{255}$Cm   &  $9.067\times10^{-6}$ \\\hline
$^{112}$Sn+$^{238}$U, $E_{\rm c.m.}=550$ MeV & & DNS Cal. \\
 &  $^{220}$Np   &  $4.046\times10^{-5}$ \\
 &  $^{221}$Np   &  $4.115\times10^{-5}$ \\
 &  $^{222}$Np   &  $1.032\times10^{-3}$ \\
 &  $^{223}$Np   &  $9.612\times10^{-4}$ \\
 &  $^{224}$Np   &  $2.321\times10^{-3}$ \\
 &  $^{222}$Pu   &  $6.793\times10^{-6}$ \\
 &  $^{223}$Pu   &  $5.258\times10^{-5}$ \\
 &  $^{224}$Pu   &  $7.271\times10^{-5}$ \\
 &  $^{225}$Pu   &  $2.902\times10^{-4}$ \\
 &  $^{226}$Pu   &  $2.899\times10^{-4}$ \\
 &  $^{227}$Pu   &  $1.222\times10^{-3}$ \\
 &  $^{226}$Am   &  $8.397\times10^{-6}$ \\
 &  $^{227}$Am   &  $1.503\times10^{-5}$ \\
 &  $^{228}$Am   &  $7.896\times10^{-5}$ \\
 &  $^{229}$Am   &  $8.820\times10^{-5}$ \\
 &  $^{229}$Cm   &  $7.720\times10^{-6}$ \\
 &  $^{230}$Cm   &  $9.225\times10^{-6}$ \\
 &  $^{231}$Cm   &  $5.181\times10^{-4}$ \\
 &  $^{232}$Cm   &  $9.033\times10^{-4}$ \\
\hline
\end{tabular*}
}
\renewcommand{\arraystretch}{1}  %%%设置表格行高
\vspace*{-6mm}
\end{table}

\newpage
\setcounter{table}{0}
\begin{table}[h!]
\doublerulesep 0.3pt
\tabcolsep 7.8mm
\setlength{\tabcolsep}{4pt}   %%%设置表格平均分配列间距
\renewcommand{\arraystretch}{1.29}  %%%设置表格行高
\caption{\hfill (continued)}
\vspace*{-1mm}
\footnotesize{\begin{tabular*}{86mm}{ccc}\hline
\parbox[t]{42.5mm}{\centering Reactions} & Isotopes & Cross section $(\textrm{mb})$\\\hline\hline
 &  $^{232}$Bk   &  $2.169\times10^{-5}$ \\
 &  $^{233}$Bk   &  $6.574\times10^{-5}$ \\
\parbox[t]{42.5mm}{\centering $^{136}$Xe+$^{198}$Pt, $E_{\rm c.m.}=643$ MeV}  & & ImQMD Cal.\\
 &  $^{156}$Pr   &  $7.000\times10^{-3}$ \\
 &  $^{157}$Pr   &  $1.475\times10^{-3}$ \\
 &  $^{158}$Pr   &  $1.435\times10^{-3}$ \\
 &  $^{159}$Pr   &  $1.427\times10^{-2}$ \\
 &  $^{160}$Pr   &  $5.893\times10^{-3}$ \\
 &  $^{162}$Nd   &  $1.440\times10^{-6}$ \\
 &  $^{166}$Sm   &  $1.701\times10^{-6}$ \\
 &  $^{169}$Eu   &  $3.668\times10^{-3}$ \\
 &  $^{170}$Eu   &  $1.333\times10^{-3}$ \\
 &  $^{171}$Eu   &  $7.671\times10^{-4}$ \\
 &  $^{172}$Gd   &  $1.964\times10^{-3}$ \\
 &  $^{173}$Tb   &  $7.281\times10^{-3}$ \\
 &  $^{174}$Tb   &  $1.950\times10^{-3}$ \\
 &  $^{175}$Tb   &  $1.375\times10^{-5}$ \\
 &  $^{175}$Dy   &  $2.417\times10^{-3}$ \\
 &  $^{176}$Dy   &  $5.629\times10^{-6}$ \\
 &  $^{177}$Dy   &  $1.701\times10^{-3}$ \\
 &  $^{178}$Dy   &  $1.960\times10^{-3}$ \\
 &  $^{179}$Dy   &  $3.928\times10^{-6}$ \\
 &  $^{177}$Ho   &  $7.389\times10^{-3}$ \\
 &  $^{178}$Ho   &  $3.530\times10^{-4}$ \\
 &  $^{179}$Ho   &  $5.104\times10^{-6}$ \\
 &  $^{179}$Er   &  $1.108\times10^{-2}$ \\
 &  $^{180}$Er   &  $9.413\times10^{-5}$ \\
 &  $^{181}$Er   &  $2.218\times10^{-3}$ \\
 &  $^{182}$Er   &  $1.060\times10^{-5}$ \\
 &  $^{182}$Tm   &  $4.233\times10^{-3}$ \\
 &  $^{183}$Tm   &  $6.283\times10^{-6}$ \\
 &  $^{184}$Tm   &  $2.488\times10^{-3}$ \\
 &  $^{186}$Yb   &  $2.035\times10^{-3}$ \\
 &  $^{189}$Lu   &  $3.665\times10^{-6}$ \\
 &  $^{191}$Hf   &  $5.717\times10^{-3}$ \\
 &  $^{192}$Hf   &  $8.504\times10^{-3}$ \\
 &  $^{193}$Hf   &  $3.355\times10^{-3}$ \\
 &  $^{194}$Hf   &  $4.254\times10^{-5}$ \\
 &  $^{200}$Re   &  $1.019\times10^{-2}$ \\
 &  $^{201}$Re   &  $4.785\times10^{-4}$ \\
 &  $^{202}$Re   &  $1.479\times10^{-3}$ \\
 &  $^{203}$Re   &  $2.356\times10^{-5}$ \\
 &  $^{204}$Os   &  $2.691\times10^{-3}$ \\
 &  $^{205}$Os   &  $9.385\times10^{-4}$ \\
 &  $^{207}$Os   &  $3.534\times10^{-5}$ \\
 &  $^{207}$Ir   &  $2.488\times10^{-3}$ \\
 &  $^{209}$Pt   &  $3.545\times10^{-3}$ \\
 &  $^{211}$Pt   &  $3.522\times10^{-4}$ \\
 &  $^{211}$Au   &  $1.375\times10^{-5}$ \\
 &  $^{221}$Pb   &  $3.915\times10^{-4}$ \\
 &  $^{222}$Pb   &  $1.178\times10^{-6}$ \\
\hline
\end{tabular*}
}
\renewcommand{\arraystretch}{1}  %%%设置表格行高
\vspace*{-6mm}
\end{table}

In addition, radioactive ion beam facilities can provide very neutron-rich projectiles.\ Some facilities,
such as the high-intensity heavy-ion accelerator facility (HIAF) in China, have a plan to produce new heavy
neutron-rich nuclei with intense secondary beams of exotic radioactive nuclei by the MNT reactions.\ The production
mechanism of heavy neutron-rich nuclei is also investigated by using the NNT reactions with radioactive
projectiles.\ We find that neutron-rich radioactive beams can improve significantly the production cross
sections of neutron-rich nuclei, see Refs.~\cite{ref2,ref5,XeW,pt1}.

\vspace*{1.8mm}
\section{Conclusions and perspectives}
\vspace*{1.2mm}

\noindent
The multinucleon transfer process will be one promising approach for producing neutron-rich
light nuclei, neutron-rich heavy nuclei, neutron-rich superheavy nuclei, and neutron-deficient
heavy nuclei.

For producing unknown nuclei in multinucleon transfer reactions, the main difficulties
in the experiments would be how to separate a given nucleus from other transfer products.\ Coupling large
gamma arrays to the new generation large solid angle spectrometers can give a full identification of
reaction products. In the future, with the increasing efficiency of experimental process,
the products with very low production cross sections can be detected.

The multinucleon transfer reactions with stable combinations can be the candidates for
producing unknown neutron-rich nuclei. However, the production cross section decreases strongly
with the increasing neutron number of objective nuclei. Therefore, the production of neutron-rich
nuclei far from the stability line with stable beam induced transfer reaction is very hard.\ Radioactive
beam induced reactions could enhance the cross sections strongly. In present experimental facilities,
the beam intensities of radioactive beams are still very low. With the development of modern radioactive
equipment, the radioactive beams could be favorable for producing neutron-rich nuclei far from the stable line.

The multinucleon transfer process is promising for producing neutron-rich SHN,
which can reach the island of stability. Theoretical predictions show that within the collisions
of actinide nuclei, the trans-actinide nuclei can be produced. And the yields decrease strongly
with the increasing charge number of objective SHN.\ It seems like that it is almost impossible
to produce SHN with $Z=114$ or larger. Recently, experimental observation of alpha decay energies
reaching as high as 12 MeV suggests the production of high atomic number, which implies that the production
cross sections of SHN through multinucleon transfer reactions were underestimated by theoretical models.\ Experimentally,
the improvement of catcher array with high granularity, better energy resolution, and linear energy response can search
the SHN with very low production rate. Theoretically, according to the experimental results, one should make more
accurate predictions of fission barrier in SHN region, which strongly influence the survival probability of
products.\ Furthermore, the shell effects, dynamical process of fission, and beta decay feeding by the
neighboring nuclei should also be investigated for better predictions of production rates.

Until now, the observed products in multinucleon transfer reactions are around the projectile and
the target.\ For producing neutron-rich superheavy nuclei, such as approaching the island of stability,
plenty of nucleons should be transferred. Due to lack of experimental data, the prediction ability of
models is limited for production of neutron-rich superheavy nuclei, although most of present
theoretical models can show good descriptions of available experimental data. Therefore,
more experiments are necessary of producing trans-actinide nuclei in multinucleon transfer
reactions, which could provide information for constraining the models. For development of
the macroscopic models, appropriate collective degrees of freedom should be considered.\ For example,
during the deep inelastic collisions, the collective vibration probably plays an important role on
charge equilibration process. Also, because of oscillations in collision process, bremsstrahlung
approach would emit photons, which could lower the effective incident energy. The double nucleon
transfer is very important in transfer channels.\ For most of theoretical models, the description
of paring mode is lack, which limits the abilities of prediction. Therefore, the improvement of
theoretical framework from the structure point is necessary.

\vspace*{-0.5mm}

\acknowledgements{We warmly thank Prof. A. Arima for the encouragement on transport model on heavy
ion collisions. F. S. Zhang thanks Y. Abe, S. Ayik, D. Boilley, L. W. Chen, D. Q. Fang, Z. G. Gan,
S. Heinz, Y. G. Ma, J. Meng, B. A. Li, C. J. Lin, W. P. Liu, Z. Liu, W. Loveland, Z. Z. Ren,
W. Q. Shen, X. D. Tang, B. Tsang, Yu. S. Tsyganov, M. Veselsky, J. S. Wang, C. Y. Wong,
G. Q. Xiao, Z. G. Xiao, F. R. Xu, Y. L. Ye, W. L. Zhan, H. Q. Zhang, Y. H. Zhang, E. G. Zhao, Y. M. Zhao,
S. G. Zhou, X. H. Zhou, and S. Zhu for valuable discussions on multinucleon transfer reactions.\ This
work was supported by the National Natural Science Foundation of China under Grants Nos.~11635003,
11025524, 11161130520, 11605270, 11605296, 11805015, and 11805280; the National Basic Research Program of
China under Grant No.~2010CB832903; the European Commission's 7th Framework Programme
(Fp7-PEOPLE-2010-IRSES) under Grant Agreement Project No.~269131; the China Postdoctoral
Science Foundation (Grant Nos.~2016M600956, 2017M621035, and 2018T110069); the Beijing Postdoctoral
Research Foundation (2017-zz-076); and the Natural Science Foundation of Guangdong Province,
China (Grant No.~2016A030310208).}

%the Natural Science Foundation of Guangdong Province, China (Grant No. 2016A030310208).
%\bibliographystyle{bnu-phd-thesis}
%\clearpage
%\bibliography{endnote}

\end{document}